\PassOptionsToPackage{colorlinks=true, linkcolor=blue, citecolor=blue, urlcolor=blue, bookmarks=false}{hyperref}

\documentclass[%
 reprint,
superscriptaddress,
 amsmath,amssymb,
 aps,
prl,
]{revtex4-2}

\usepackage{subcaption}

\newcommand{\orcid}[1]{\href{https://orcid.org/#1}{\includegraphics[width=8pt]{orcid_ID.png}}}

\usepackage{graphicx}
\usepackage{dcolumn}
\usepackage{bm}
\usepackage{soul} 
\usepackage{amsmath}
\usepackage{newunicodechar}
\usepackage{orcidlink}

\newunicodechar{β}{\beta}

\usepackage{xcolor}

\newcommand{\dc}[1]{\textcolor{cyan}{[DC: #1]}}

\begin{document}

\preprint{APS/123-QED}

\title{Measuring Small-scale Turbulence in Supernova Remnants with Jitter Radiation}

\author{Emily Simon \orcidlink{0000-0002-9386-630X}}
 \email{ersimon@uchicago.edu}
\affiliation{%
 Department of Astronomy \& Astrophysics, University of Chicago, Chicago, IL 60637, USA}

\author{Damiano Caprioli \orcidlink{0000-0003-0939-8775}}
\affiliation{%
 Department of Astronomy \& Astrophysics, University of Chicago, Chicago, IL 60637, USA}
 \affiliation{%
 Enrico Fermi Institute, The University of Chicago, Chicago, IL 60637, USA}

 \author{Niccolò Bucciantini \orcidlink{0000-0002-8848-1392}}
\affiliation{%
 INAF Osservatorio Astrofisico di Arcetri, Largo Enrico Fermi 5, 50125 Firenze, Italy}
 \affiliation{%
 Dipartimento di Fisica e Astronomia, Università degli Studi di Firenze, Via Sansone 1, 50019, Sesto Fiorentino (FI), Italy}
 \affiliation{%
 Istituto Nazionale di Fisica Nucleare, Sezione di Firenze, Via Sansone 1, 50019, Sesto Fiorentino (FI), Italy}

\author{Emanuele Greco \orcidlink{0000-0001-5792-0690}}
 \affiliation{%
 INAF-Osservatorio Astronomico di Palermo, Piazza del Parlamento 1, 90134, Palermo, Italy}

\author{Marco Miceli \orcidlink{0000-0003-0876-8391}}
\affiliation{%
Dip. di Fisica e Chimica E. Segrè, Università degli Studi di Palermo, Piazza del Parlamento 1, 90134 Palermo, Italy}
\affiliation{%
INAF-Osservatorio Astronomico di Palermo, Piazza del Parlamento 1, 90134, Palermo, Italy}




\date{\today}

\begin{abstract}
Hard X-ray emission from Cas A has been attributed to jitter radiation, implying magnetic turbulence on scales near or below the ion Larmor radius. We propose these fields arise from ion-scale micro-instabilities (e.g., mirror and firehose) in the shock downstream, and use high-resolution hybrid simulations to demonstrate the generation of such small-scale turbulence. Its spectral index, once numerical dissipation is accounted for, is consistent with that inferred for Cas A, suggesting that jitter radiation is a direct probe of ion-scale turbulence in supernova remnants. 
\end{abstract}

\maketitle


\begin{figure*}
    \centering
    \includegraphics[width=0.45\linewidth]{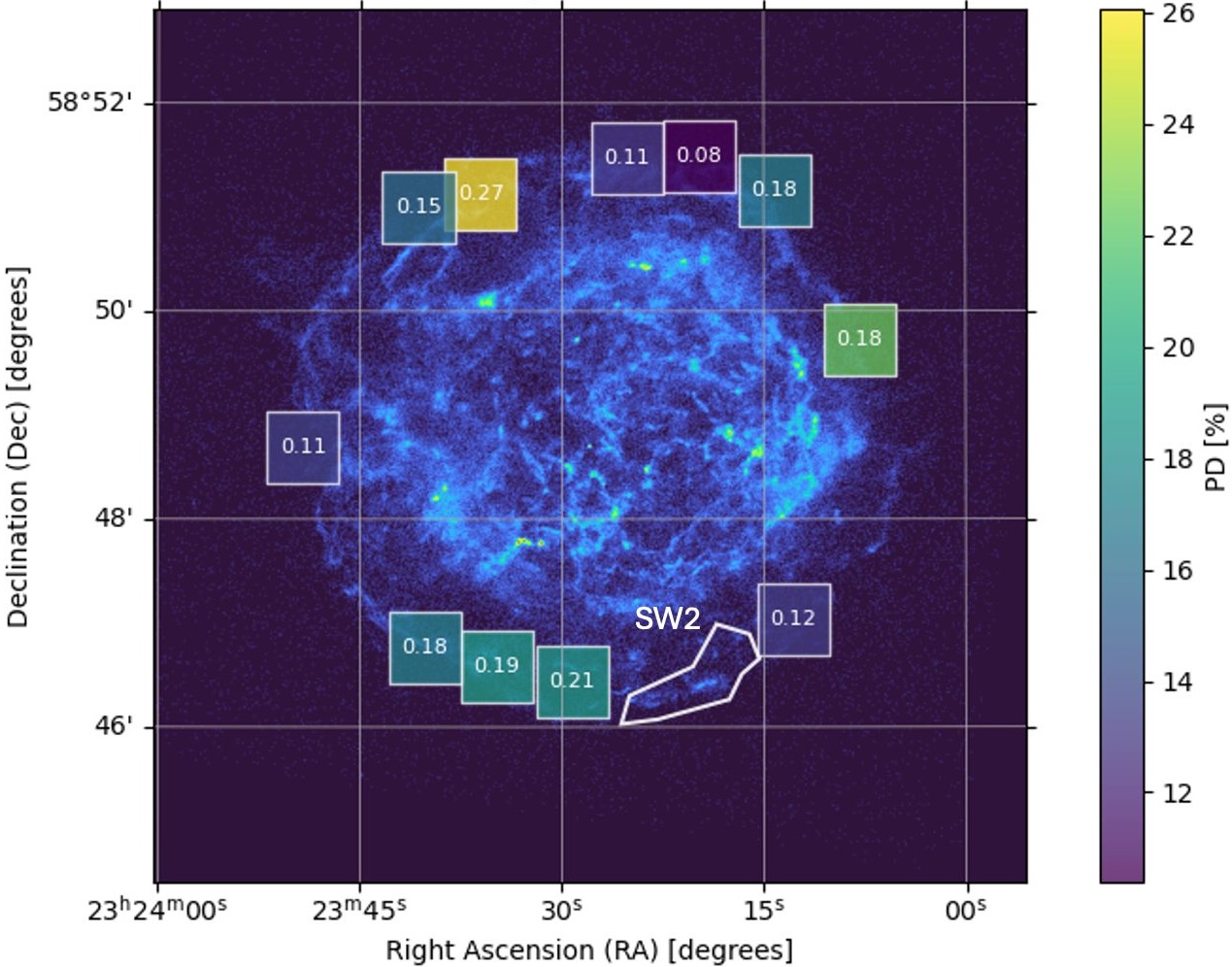}
    \includegraphics[width=0.45\linewidth]{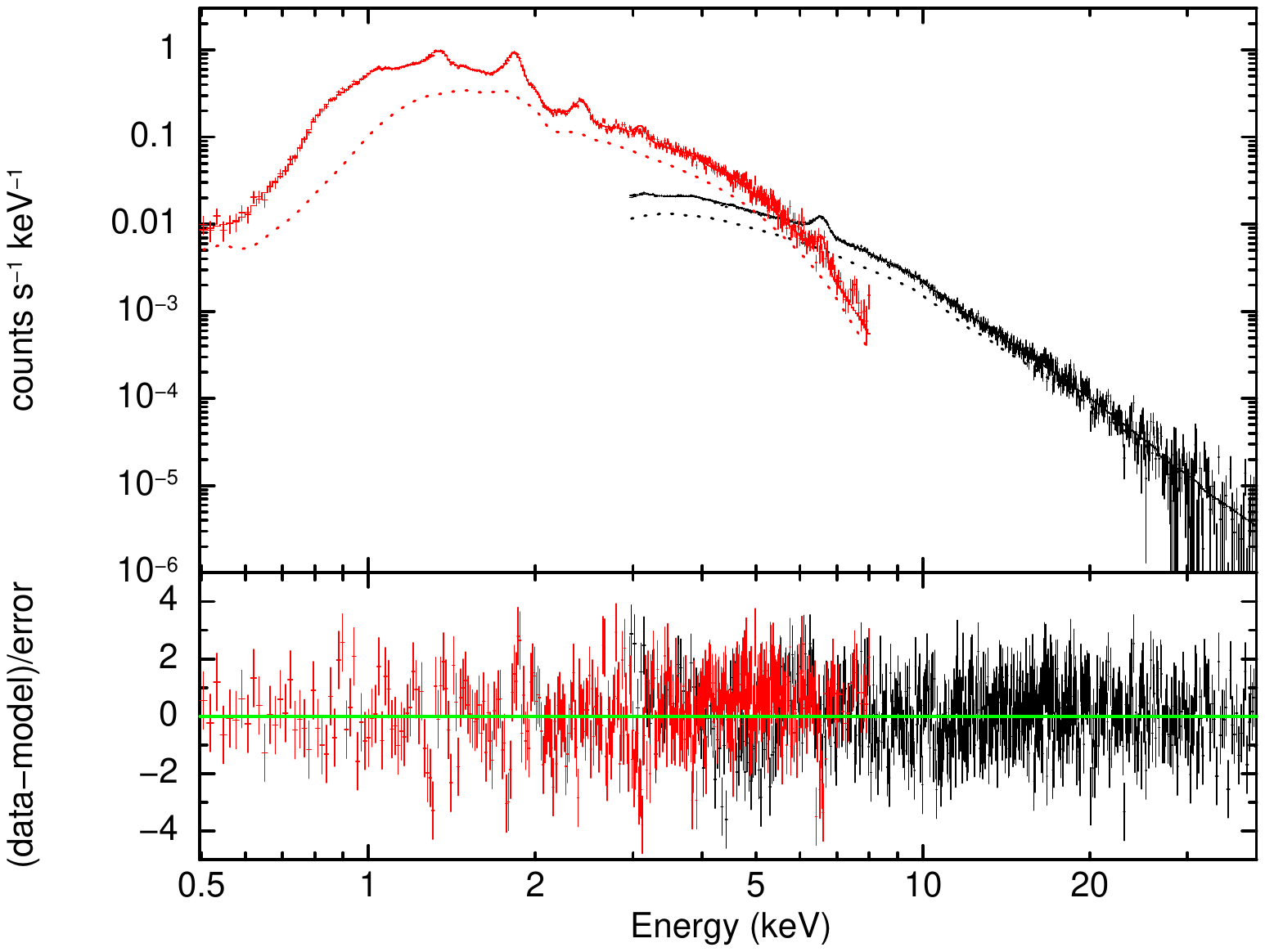}
    \caption{\textit{Left panel.} Chandra image of Cas A in the 4-6 keV band, overlaid with boxes from \cite{mgv25} showing the polarization degree (PD) in each area, courtesy of A. Mercuri and with the white polygon marking the SW2 region analyzed in this paper. \textit{Right panel.} Chandra (red points) and NuSTAR (black points) spectra of the SW2 region. The solid lines represent the total best-fit model, the dotted lines represent the power-law contribution only. Lower panel shows the residuals.}
    \label{fig:observations}
\end{figure*}

\section{Introduction}
\label{Introduction}
Probing the properties of turbulence in collisionless  plasmas is one of the most prominent open question in astrophysics.
The standard model for turbulence in the interstellar medium (ISM) is a Kolmogorov or Iroshnikov--Kraichnan/Goldreich--Shridar cascade \cite{kolmogorov41, iroshnikov64, goldreich+95}, where energy is injected at large ($10-100$ pc) scales by macroscopic astrophysical phenomena (e.g., supernova explosions, galactic density waves) and cascades down to smaller scales.
The subsequent spectrum, which scales as $P(k) \propto k^{-5/3}$ in the case of the observed Kolmogorov turbulence \citep[e.g.,][]{armstrong+95}, is expected to terminate near the ion Larmor radius ($\sim 100$ km), where energy is dissipated into heat via kinetic effects \cite{lithwick+01, schekochihin+09}.
Therefore, a turbulence cascade over more than 10 decades in wavelength should have negligible power at the scales of the thermal ion Larmor radius, which in the ISM is also comparable to the ion inertial length.

A recent work analyzing combined X-ray observations of the supernova remnant (SNR) Cassiopeia A (Cas A) \cite{greco+23} shows a power-law emission extending beyond the synchrotron cutoff with a photon index $\Gamma \sim 3 -3.5$, which was suggested to come from jitter radiation.
Synchrotron arises when relativistic electrons gyrate in a field whose coherence scale $\lambda_c$ exceeds the emission arc-length $l$---the arc over which the $1/\gamma$ beaming cone sweeps past the observer, $l = r_{\rm L}/\gamma\ll \lambda_c$, where $r_L=\gamma m_ec^2/(eB)$ is the electron Lorentz factor.
Jitter, instead, is produced in the opposite limit in which there is power at scales smaller than $r_L$, i.e., $\lambda_c \ll m_ec^2/(eB)$ \citep{toptygin+87, medvedev00, kelner+13}, independent of the electron's Lorentz factor.

Assuming a magnetic field in the shock downstream of Cas A with strength $\sim 100 \mu \rm{G}$, the presence of jitter radiation implies that there must be large-amplitude turbulence ($\delta B/B_0 \sim 1$) on the scale $\lambda_c \lesssim 170  \, \rm{km}$. 
The visibility of jitter radiation is determined by its emissivity relative to synchrotron, which scales as $S^{\rm jit}/S^{\rm syn} \sim (\delta B/B_0)^2$. 
Additionally, for a turbulent spectrum $\propto k^{-\nu_B}$, the jitter radiation returns a power-law spectrum with $\Gamma = \nu_B+1$, which, unlike synchrotron, does not depend on the electron spectrum \citep{reville+10, mao+11}; we derive this relation and the dependence of the jitter-to-synchrotron ratio on $\delta B/B_0$ in the End Matter. 
Therefore, the measurement of jitter in Cas A \cite{greco+23} constrains both normalization and spectrum of the post-shock magnetic turbulence at microphysical scales of the order of $\sim 100$ km, comparable to the ion inertial length  $d_i\approx  228 \sqrt{n}\, \rm{km}$, where $n$ is the plasma number density in units of 1 proton $\rm{cm}^{-3}$.

Given that in the ISM $\lambda_c\sim 10-100$ pc, the fluctuations responsible for jitter cannot come from extrinsic ISM turbulence, which prompts the search for an intrinsic (self-generated) origin.
Cosmic-ray (CR) streaming is a possible candidate and in young SNRs is expected to proceed via the Bell instability \citep{bell04, amato+09}.
For typical CR currents, its growth is fastest at a wavelength $k_{\rm max}^{-1} \approx  10^{9} \, \rm{km}$ \citep{bell+13}, with power moving at even larger scales during the nonlinear stages towards saturation \cite{zacharegkas+24}, still orders of magnitude above jitter-inducing ones.

We suggest that jitter may be due to plasma\textit{ microinstabilities}, particularly firehose and mirror, which arise from pressure anisotropies when magnetic tension is low compared to transverse (firehose) or compressive (mirror) perturbations  \citep{rosenbluth56, vedenov+58, chandra+58, parker+58}.
Mirror and firehose are known to produce turbulence on $d_i$ scales in weakly magnetized plasmas with thermal/magnetic pressure ratio $\beta\gg1$ \citep{kunz+14b, schekochihin+08}, and strong shocks preferentially convert their free kinetic energy into heat rather than magnetic fields, even when CR instabilities are effective \citep{haggerty+20}.

In this work, we validate this hypothesis via hybrid (fluid electrons--kinetic ions) plasma simulations of strong SNR shocks, demonstrating that micro-instabilities naturally inject magnetic energy at the kinetic scales necessary to produce jitter radiation consistent with X-ray observations of Cas A.
These results represent both an unprecedented probe of magnetic turbulence at small kinetic scales and the evidence of the prominence of microinstabilities in collisionless astrophysical plasmas.

\section{Jitter-dominated regions in Cas A}
\label{data_analysis}
Being sensitive to small-scale micro-instabilities, jitter radiation is expected to be intrinsically unpolarized, whereas synchrotron can be polarized up to $\sim 80 \%$, depending on the photon index $\Gamma$ \cite{ginzburg-syrovatskii68}. A lack of observed polarization is therefore a necessary, though not sufficient, condition for jitter radiation. In fact, projection effects and large-scale turbulence, such as the Bell instability, can also mitigate or even cancel the observed polarization. The maximum X-ray polarization of $\sim 40\%$ across all the SNRs observed by IXPE \citep{IXPE} so far was found in the southwestern limb of SN 1006 by \cite{zsp25} while typical values elsewhere range between 10\% and 20$\%$ (see \cite{sla24,greco25} for a review), far from the intrinsic limit. On the other hand, regions with zero polarization are the most promising for detecting signatures of jitter radiation

\begin{figure*}[t]
    \centering
    \includegraphics[trim= 105 20 150 4, clip=true, width=0.98\linewidth]{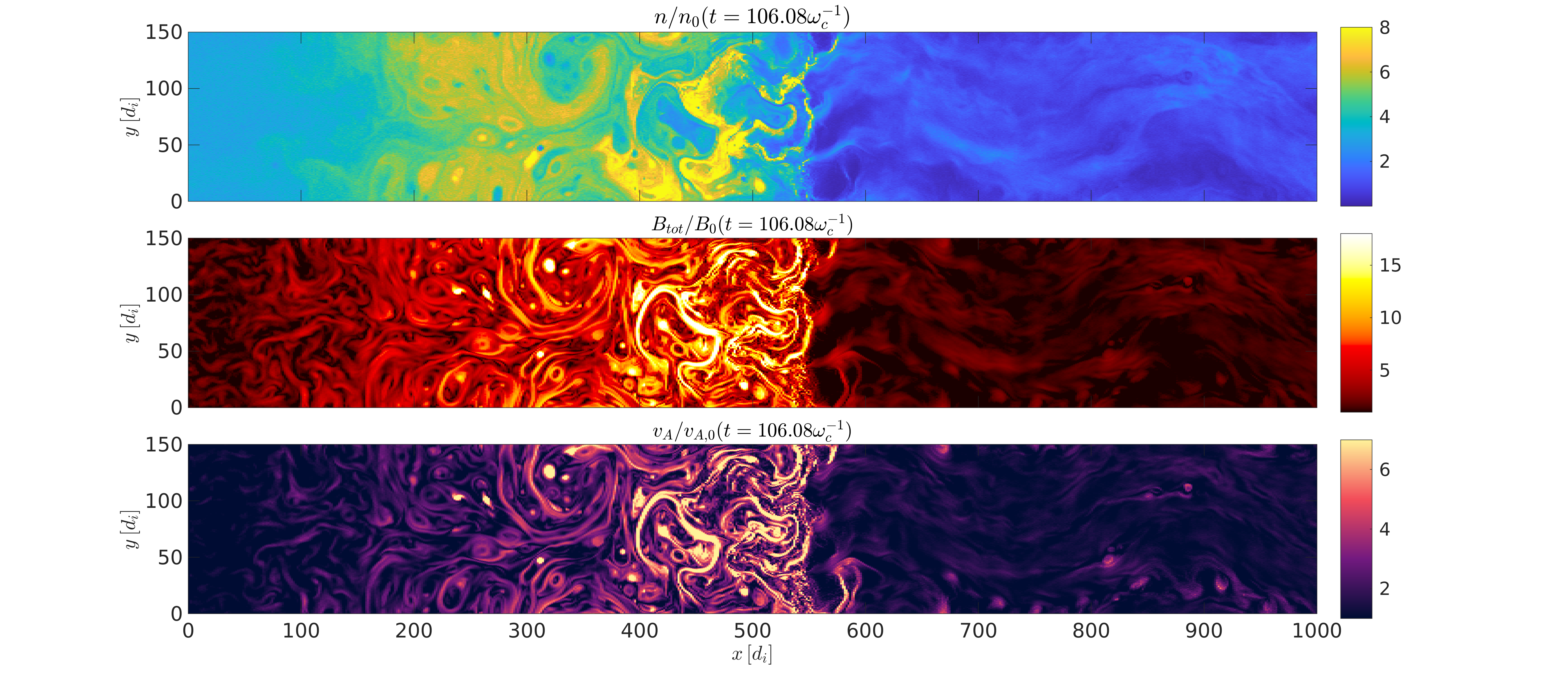}
    \caption{Snapshots from a hybrid simulation of a $M_A=20$ parallel shock. From top to bottom: plasma number density $n$, total magnetic field $B_{\rm{tot}}$, and local Alfv\'en speed $v_A$ normalized their initial values at $t=106.08 \omega_c^{-1}$.
    }
    \label{fig:benchmark_sim}
\end{figure*}

\cite{mgv25} identified regions in Cas A where the polarization fraction of the X-ray synchrotron emission ranges from 10\% to 26\%, as well as areas with zero polarization. We took advantage of these results to expand on the analysis previously performed by \cite{greco+23} by extracting Chandra/ACIS and NuSTAR/FPMA,B spectra from an extended region showing zero X-ray polarization. Fig. \ref{fig:observations} shows the Chandra Cas A image in the 4-6 keV overlaid with the polarization degree values from \cite{mgv25} and with the new SouthWest2 (SW2) shock region.

As a first step, we followed the same approach as \cite{greco+23} by fitting the NuSTAR spectra between 4 and 40 keV. A proper description of the spectra is achieved only when adopting a model with no hard cutoff, i.e. jitter or power-law component. In particular, when adopting the \texttt{zira} model by \cite{za07} (a loss-limited model of synchrotron emission), the \texttt{SRcut} model by \cite{rk99} (synchrotron radiation from an energy spectrum of electrons following a power law with exponential cutoff) and a straight power-law, we obtained $\chi^2$ values of 527, 234 and 124, respectively, with 105 degrees of freedom (d.o.f.).  By adopting a jitter radiation model instead of a straight power-law we do not have a significant improvement in the description of the spectra, indicating that the synchrotron-to-jitter transion $\omega_{sj}$ (End Matter) lies below the NuSTAR band ($\leq 4$ keV), and the $4-40$ keV emission is jitter-dominated. 

Therefore, we also included in our analysis the Chandra data and the soft emission between 0.5 and 4 keV, which is dominated by thermal plasma, now covering almost two decades in energy and enabling a more robust constraint on the shape of the nonthermal component. A satisfactory description of the thermal emission was reached when adopting a model with two \texttt{vpshock} thermal components, on top of the nonthermal one. Discussing the features of the thermal emission is beyond the scope of this work, so we limit to highlight that we find the thermal plasma to be out of equilibrium of ionization and with solar or slightly supersolar abundances, a configuration typical of freshly-shocked circumstellar medium \cite{vas25}.

In Fig. \ref{fig:observations} Chandra and NuSTAR spectra of region SW2 are shown with corresponding residuals for the model with two \texttt{vpshock} components and a straight power-law. The $\chi^2/$d.o.f. values corresponding to this setup for region SW2 is 1245/944, against 1320/944 and 1389/944 for the scenarios with \texttt{srcut} and \texttt{zira} components replacing the power-law, respectively. We retrieved a best-fit photon index $\Gamma = 3.12_{-0.07}^{+0.08}$, translating into a turbulence spectral index of $\nu_{\rm{B}} =  \Gamma - 1 = 2.12_{-0.07}^{+0.08} $.

\section{Simulation Results}

We use the {\tt dHybridR} code \citep{haggerty+19a} to run a high-resolution, 2D simulation of a non-relativistic shock with Alfv\'enic Mach number $M_A = v_{\rm sh}/v_A = 20$, where $v_{\rm sh}$ $v_A$ are the shock and Alfv\'en speeds, propagating in a $\beta=1$ medium (so that also the sonic Mach number $\sim M_A$);
the setup is discussed extensively in \cite{haggerty+20}, with further details in End Matter.
Figure \ref{fig:benchmark_sim} shows a snapshot at time $t=106.08 \omega_{c}^{-1}$ (where $\omega_{c}^{-1}$ is the ion cyclotron time) of plasma density, total magnetic field strength, and Aflv\'en speed, all normalized to their intial values (top to bottom, respectively); 
at this time, the shock is at $x\sim 540 \, d_i$.
Upstream fluctuations are entirely seeded by back-streaming CRs via the Bell instability \citep{reville+13, caprioli+14b}.

\begin{figure*}
    \centering
    \includegraphics[trim= 16 2 12 0, clip=true, width=0.99\linewidth]{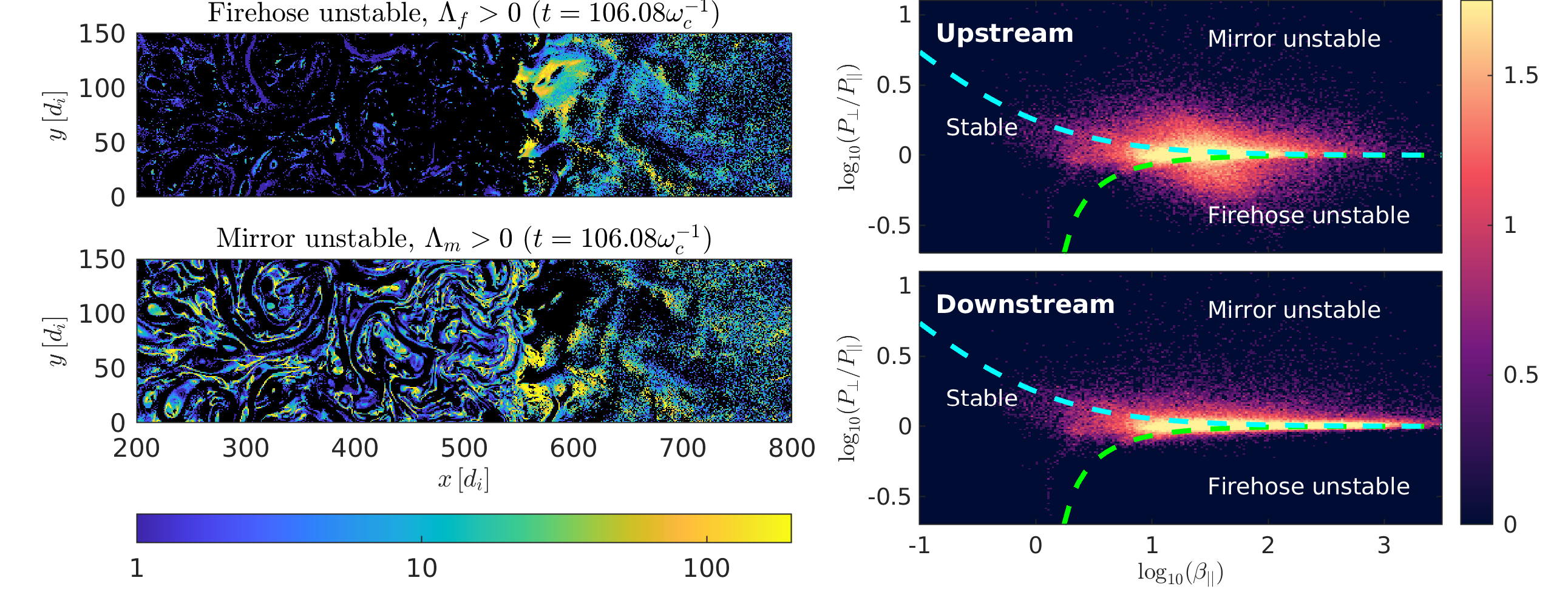}
    \caption{\textit{Left:} Smoothed regions which are firehose (top) or mirror (bottom) unstable in the downstream ($x<540\, d_i$) and upstream. The colorbar shows the local value of $\Lambda_{f,m}$ respectively. \textit{Right:} 2D histograms of the pressure anisotropy as a function of plasma $\beta_\parallel$ for both upstream (top) and downstream (bottom) regions. The colorbar represents the log of the number of $1 \, d_i^2$ sections which reside in a given bin.}
    \label{fig:fm_brazil_combined}
\end{figure*}

Microinstabilities arise from pressure anisotropies and are more easily triggered in high-$\beta$ plasmas.
In linear theory \cite{chandra+58, barnes66, pokhotelov+04} and in simulations \cite{kunz+14b}, the instability thresholds for firehose ($\Lambda_f$) and  mirror ($\Lambda_m$) depend on the pressure anisotropy as $\Lambda_f = P_\parallel - P_\perp - 2P_B > 0$ and $\Lambda_m = P_\perp - P_\parallel -\frac{P_\parallel}{2 P_\perp} 2 P_B> 0$, where $\perp$ and $\parallel$ are taken with respect to the local B-field orientation.


Figure \ref{fig:fm_brazil_combined} (left) shows the values of  $\Lambda_{f,m}$, smoothed into $1 \, d_i^2$ patches to reduce numerical noise and highlight coherent structures, for our benchmark shock run. Self-generated fluctuations induce strong departures from marginal stability ($\Lambda_{f,m} \sim \mathcal{O}(100)$) in various regions, especially upstream. 
The right panel in Figure \ref{fig:fm_brazil_combined} shows the 2D histogram of the thermal pressure anisotropy as a function of the plasma $\beta_\parallel$ in two regions $540\, d_i$ wide both upstream (top) and downstream (bottom).
The cyan/green contours mark the boundaries above/below which the plasma is respectively mirror/firehose unstable \citep{hellinger+06}. 
Note how the many unstable upstream regions collapse at or below marginal stability in the downstream, with the trend being more evident for firehose.


Such a rapid growth and saturation of the firehose instability is consistent with linear theory, $\gamma_f/\omega_c \simeq (k_{\parallel, \perp} d_i) \sqrt{\beta_i} \left( \Lambda_f /2\right)^{1/2}  \gtrsim 1$ for $\beta\gg1$ and given that the wavenumber of the fastest-growing mode in either the parallel or perpendicular directions $k_{\parallel, \perp}\sim 1/d_i$  \citep{gary+98, hellinger+00, schekochihin+08}.
At the same time, the linear theory suggest that the mirror instability is much slower growing \citep{pokhotelov+04, kunz+14b}, and thus mirror-unstable regions persist farther into the downstream. 
The free energy in the pressure anisotropy is converted into amplified magnetic fields, consistent with the strong enhancements observed immediately behind the shock where firehose is likely to be acting.
The time needed for mirror to unravel may exceed the duration of the simulation, and thus may be expected to contribute to the turbulence spectrum further downstream, also possibly fed by firehose-seeded fluctuations.

We stress that theoretical expectations are derived for small perturbations beyond marginal stability \citep{hellinger+00, kunz+14b}, thus do not apply directly to the large anisotropies $\Lambda_{m,f}\gg1$ impulsively produced by a strong shock. 
A quantitative treatment of the growth and saturation is beyond the scope of this Letter, but the extrapolation of the linear theory qualitatively supports the idea that microinstabilities are strongly triggered at shocks that accelerate CRs, and their nonlinear evolution leads to prominent magnetic field amplification.


\section{The spectrum of turbulence on small scales}

\begin{figure}
    \centering
    \includegraphics[trim= 6 6 4 5, clip=true, width=0.98\linewidth]{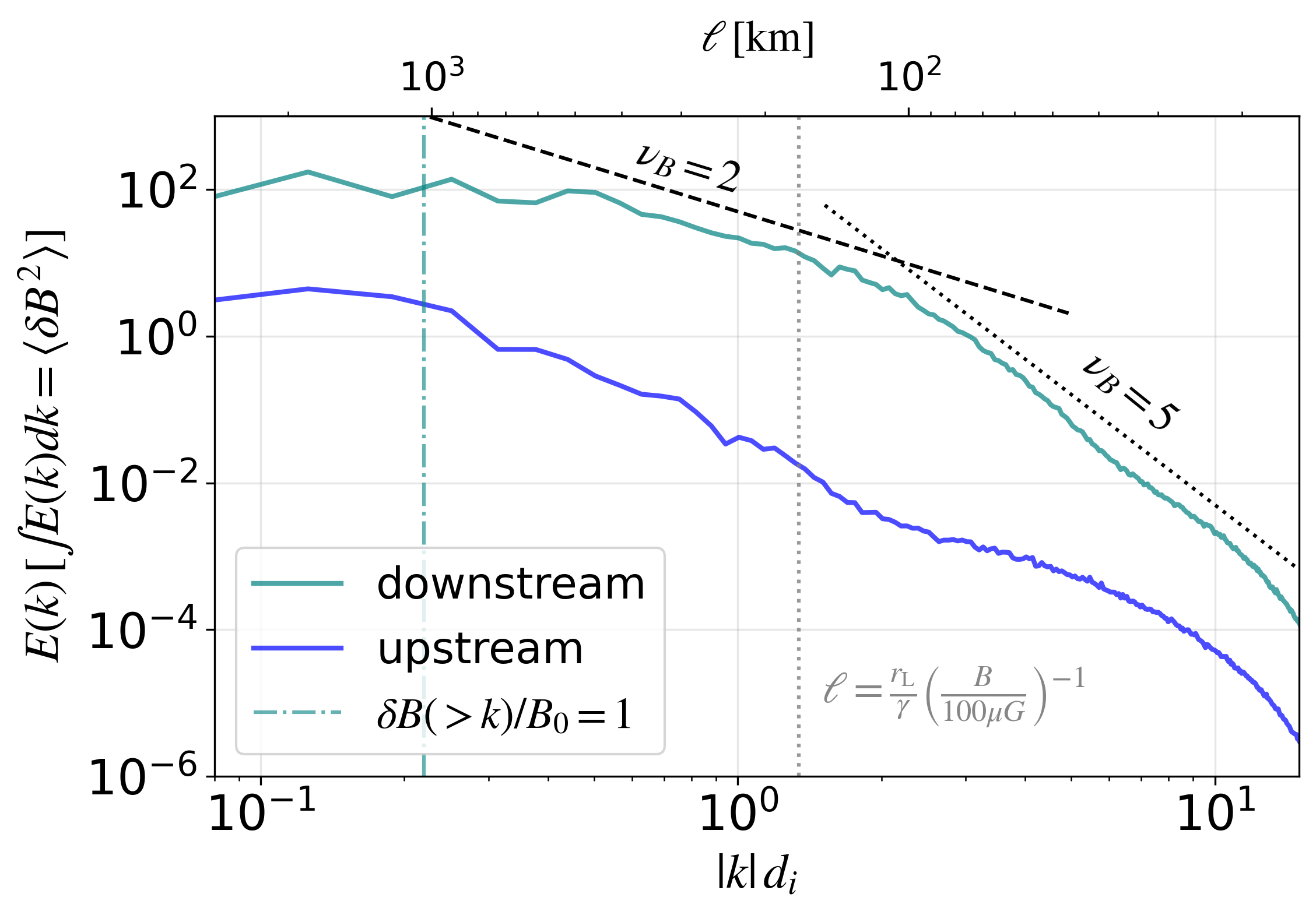}
    \caption{$k$-shell-averaged magnetic energy spectrum $E(k)$, calculated downstream ($x = 450- 550 \, d_i$, teal) and upstream ($560-660 \,d_i$, blue). 
   The teal vertical line represents the wavenumber where the cumulative power at all larger $k$ becomes equivalent to the mean downstream field. 
The upper axis shows the length scale of the turbulence, $\ell = 1/k \, \rm{km}$, assuming $d_i = 228 \, \rm{km}$, and the gray vertical line marks the reference scale where $\ell$ equals the electron Larmor radius for $B = 100 \mu \rm{G}$.}
    \label{fig:fourier}
\end{figure}

We now address the properties of the small-scale turbulence triggered by  microinstabilities. 
Figure \ref{fig:fourier} shows the $k$-shell-averaged magnetic energy spectrum, $E(k) = \int_{k}^{\infty} E(k)\,dk = \langle (\delta B/B_0)^2\rangle$, computed by taking the 2D spatial Fourier transforms of all three mean-subtracted field components and summing their squared magnitudes in a slab of thickness $100 \, d_i$ both in the downstream (teal) and upstream (blue) of the shock at the timestep $t=106.08 \, \omega_c^{-1}$.

The downstream spectrum traces $E(k)\propto k^{-\nu_B}$ with $\nu_B\approx2$ over $kd_i\lesssim2-3$, comparable to the turbulence index inferred for Cas A from its hard X-ray emission (Fig. \ref{fig:observations}). 
Beyond $k\,d_i\sim 3$ the spectrum steepens sharply ($\nu_B\approx5$), more than predicted by typical kinetic cascade or damping mechanism \citep{howes+08, boldyrev+12, passot+15, alexandrova+12}.
The location of this spectral break tracks the code's finite spatial resolution rather than a change of physics (see End Matter for  details).

The vertical, dot-dashed line marks $k^{\star}\,d_i\approx0.22$, where the cumulative fluctuation power reaches equipartition, $\int_{k^{\star}}^{\infty} E(k)\,dk = B_{\rm RMS}^{2}$, with $B_{\rm RMS}\approx8\,B_0$ the rms field strength $100\,d_i$ behind the shock;
this places most of the resolved magnetic power at physical scales of $\sim$few-$d_i$.

The upstream spectrum is comparatively smooth with spectral index $\nu_B \approx 3$ and less power at small-scales, which suggests that the shock transition generally amplifies the magnetic spectrum, possibly also through processes that locally include ion-Weibel and Richtmyer--Meshkov instabilities \citep{caprioli+13, orusa+25b}.
Though with the limitation of a finite grid resolution, the simulated magnetic spectrum supports the synchro-jitter picture in which the large-scale field acts as an effective mean field producing synchrotron, while high-$k$ power at scales comparable with the electron Larmor radius sources the jitter tail with approximately the slope observed in Cas A. 

The jitter condition can be written as $\lambda_c \ll 170 \left(B/100 \mu \rm{G} \right)^{-1} \,\rm{km}$.
Whether our simulated turbulence reaches these scales depends on density since $d_i \approx 228n^{-1/2} \, \rm{km}$; higher $n$ shrinks the ion inertial length bringing the ion-scale ($kd_i \sim 1$) turbulence we resolve down toward the $\sim 170 \, \rm{km}$ jitter scale.
For the fiducial $n = 1\,\mathrm{cm}^{-3}$, the resolved range lies above this scale, but at the higher densities inferred for Cas~A's dense shocked clumps \citep{vink+24} it reaches it.
Observationally, this may translate to spatially-varying spectra within an SNR, with jitter overtaking synchrotron at different wavenumbers in different regions.

\section{Conclusions}
While observations of SNRs such as Cas A suggest that jitter radiation may be responsible for the hard X-ray emission, standard models of extrinsic turbulence and large-scale CR-driven instabilities fail to produce the necessary magnetic power at scales of hundreds of km, comparable with the ion inertial length in the interstellar medium. 
We show that mirror and firehose micro-instabilities, driven by pressure anisotropies in the shock downstream, can supply this power: rather than waiting for energy to cascade from large scales, they ``short-circuit" the cascade, injecting magnetic energy non-locally in $k$-space directly at the scales responsible for jitter radiation (analogous to the reconnection-mediated process of \citep{boldyrev+17}). 
The spatial correlation between high-amplitude fluctuations and mirror/firehose-unstable regions indicates these instabilities are the primary drivers of the small-scale turbulence. 
The resulting spectral index agrees, over the converged range, with that inferred from an unpolarized, jitter-consistent region near the Cas A shock front, providing a coherent physical basis for the observed hard X-ray emission. 
Extending this search for jitter-dominated regions to other young SNRs will clarify how common the mechanism is.

\begin{acknowledgments}

The authors would like to thank Brian Reville and Mikhail Medvedev for helpful discussions. D.C. and E.S.~were partially supported by NASA grant 80NSSC24K0173 and NSF (grants AST-2510951 and AST-2308021).
Simulations were performed at the University of Chicago Research Computing Center.
E.G. and M.M. acknowledge support by the Italian Space Agency (Agenzia Spaziale Italiana, ASI) through contract ASI-INAF N. 2025-9-U.0 ``Unveiling magnetic turbulence in nonthermal emission of Supernova Remnants: Jitter radiation Observed with IXPE and NuSTAR Telescopes (JOINT)". E.G. acknowledges support from the INAF Minigrant RSN4 “Investigating magnetic turbulence in young Supernova Remnants through X-ray observations”.
\end{acknowledgments}


\appendix
\section{End Matter}

\subsection{Simulation Setup and Numerical Effects}
This simulation was performed in 2D and initialized such that a non-relativistic flow of thermal particles is injected from left to right with drift speed $v_d = -20 v_{A,0}$ where $v_{A,0} = B_0/\sqrt{4 \pi m_i n_i}$ is the Alfvén speed, $B_0$ is the initial magnetic field (initialized along $x$), $m_i$ is the ion mass, and $n_i$ is the ion number density.
The left edge of the box reflects the plasma, leading to a shock propagating rightward with $v_{\rm{sh}} \sim 6.5 v_A$ in the downstream frame. 
The box has dimensions $L_x \times L_y = 10^4 \times 150 \, d_i^2$, where $d_i \equiv v_A/\omega_c$ is the ion inertial length (the scale under which particles no longer perceive the magnetic field as frozen-in) and $\omega_{c} \equiv ZeB/m_i c$ is the ion gyrofrequency.  
These plasma parameters can be expressed as $d_i \approx 230 (\frac{1}{n_i} )^{1/2} \, \rm{km}$ and $\omega_{c}^{-1} \approx 3.65 (\frac{B}{3 \mu \rm{G}} )^{-1} \, \rm{minutes}$ for protons.
The resolution of the box is 10 cells per $d_i$, and the simulation is initialized with $\beta \equiv 8 \pi n_i T_i/B^2 \approx 1$ and $c=100 \, v_{\rm A}$.  


\begin{figure}[h!]
    \centering
    \includegraphics[width=0.98\linewidth]{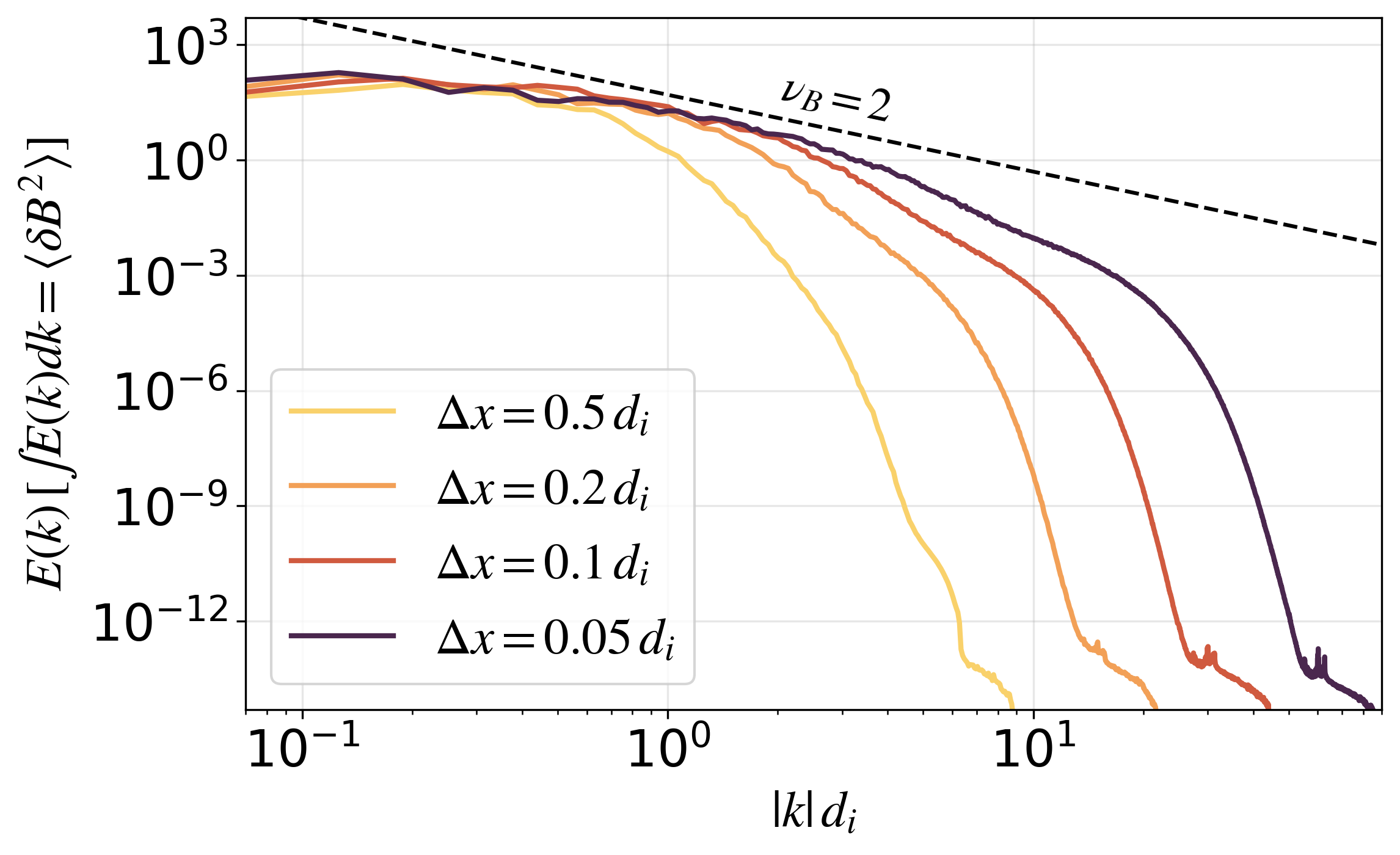}
    \caption{Comparison of downstream $k-$shell-averaged magnetic power spectrum in test simulations with varying cell-size, $\Delta x \equiv d_i/\rm{cells}$. We demonstrate that the spectral break is not converged but increases monotonically as resolution increases.}
    \label{fig:res_tester}
\end{figure}

As discussed in the main text, the magnetic turbulence most important for jitter radiation is at sub-$d_i$ scales, and this range is potentially affected by numerical resolution and the lack of electron physics in hybrid simulations. 
Figure \ref{fig:res_tester} shows a convergence test in grid resolution: the break of the $k^{-2}$ power-law portion shifts monotonically to higher $k$ as resolution increases. 
Additionally, all hybrid codes must apply grid-scale filtering to cancel Whistler modes; a convergence test showed the power in very-high-$k$ modes grew monotonically as less filtering was applied, however this is a much smaller effect at much smaller modes ($kd_i \gtrsim 10$) than that of resolution.

A kinetic-Alfv\'en-wave (KAW) cascade has been shown to emerge at sub-Larmor scales from mirror and firehose turbulence in simulations \citep{kunz+14b} and is observed in the solar wind and Earth's magnetosheath \citep{chen+17, chettri+26}, with spectral slopes in the range of $\nu_B = 7/3$ to $8/3$ when undamped \citep{howes+08, boldyrev+12}, steepening to $\sim 11/3$ when damped \citep{passot+15, alexandrova+12}. 
All are shallower than the $\nu_B \approx 5$ we recover; the true high-$k$ slope is set by electron-kinetic physics absent in our fluid-electron model, and may be shallower than our numerically-suppressed spectrum indicates.

Dimensionality may also impact the spectral index. 
Previous hybrid simulations have found the magnetic spectrum largely insensitive to 2D versus 3D \citep{franci+18}, but dimensionality does alter particle injection and acceleration in perpendicular hybrid shock simulations \citep{orusa+25b}, which may carry over to highly-turbulent parallel shocks. 
The full implications for small-scale turbulence and cascades remains an open, computationally-challenging question.

\subsection{A simple model for Synchro-Jitter}
Consider a relativistic particle moving at a speed $\boldsymbol{v}$, and with Lorentz factor $\gamma$, in a region characterized by a magnetic field with a net mean absolute value $B$ (perpendicular to $\boldsymbol{v}$) and a fluctuating component, with rms amplitude $\delta B$. 
The particle will emit toward the observer only if $\| 1-\boldsymbol{v}\cdot \boldsymbol{n} \| < \gamma^{-2}$, where $\boldsymbol{n}$ is the direction of the observer. 
As long as $\delta B / B <1$ in the region, this condition defines an emission window with length $\lambda_{\rm e} \simeq 2 mc/e  B $. While emitting toward the observer, the particle will sample only magnetic fluctuations on scales smaller than $\lambda_{\rm e}$. To these fluctuations with wave numbers $k \gtrsim k_\lambda = 2\pi/\lambda_{\rm e}$, the particle will respond by jittering. Fluctuations on larger scales can be reabsorbed into the definition of the mean magnetic field over the emission window. The particle trajectory will be a combination of a mean circular orbit with radius $\sim \gamma m c^2 /(eB)$, and a jittering component on scales $\lesssim \lambda_{\rm e}$. 
What matters for this jittering is the power in the direction of motion of the magnetic fluctuations perpendicular to $\boldsymbol{v}$. In the relativistic regime, in the emission window, one can safely assume $\boldsymbol{v} = \boldsymbol{n}$, $\| \boldsymbol{v}\|=1$, allowing one to treat the magnetic turbulence in the so called \textit{slab}-regime.\\\\
It is possible to reformulate the Jitter theory by \citet{kak13} in this slab-regime, and one finds that the Jitter emissivity per unit frequency and unit solid angle, toward the observer is:
\begin{multline*}
    \mathcal{S}^{\rm jit}_{\omega}[\gamma,\delta B,\Psi_{\rm slab}(k),\theta] = \frac{e^4}{2\pi^3 m^2c^4}\,\delta B^2
    \frac{\gamma^2(1+\gamma^4\theta^4)}{(1+\gamma^2\theta^2)^4} \\ \,\times \Psi_{\rm slab}\left(\frac{\omega(1+\gamma^2\theta^2)}{2c\gamma^2}\right)
\end{multline*}
where $\theta$ is the angle made by the observer direction $\boldsymbol{n}$ with respect to the unperturbed orbital plane, and $\Psi_{\rm slab}(k)$ is the 1D power spectrum of the transverse components of the magnetic field in the direction of the observer, and normalized such that:
\begin{align}
 \int_{0}^\infty \Psi_{\rm slab}(k) dk =1
\end{align}
On the other hand, the standard synchrotron emissivity per unit frequency and unit solid angle, due to the mean orbital motion is:

\begin{multline*}
    \mathcal{S}^{\rm syn}_\omega[\gamma,B,\theta]= \frac{3e^2\gamma^2}{4\pi^2c}\left(\frac{\omega}{\omega_c}\right)^2[1+\gamma^2\theta^2] \times
    \\
    \left[[1+\gamma^2\theta^2]K_{2/3}^2\left(\frac{\omega}{2\omega_c}\right) + \gamma^2\theta^2 K_{1/3}^2\left(\frac{\omega}{2\omega_c}\right)\right]
\end{multline*}

where:
\begin{align}
\omega_c = \frac{3 e \gamma^2 B}{2 m c}
\end{align}
One can combine the two to get:
\begin{multline*}
    \mathcal{S}_\omega^{\rm sj}[\gamma,B,\delta B, \Psi_{\rm slab}(k),\theta] = 
    \\
    \begin{cases}\mathcal{S}^{\rm syn}_\omega[\gamma,B,\theta] \quad{\rm for} \quad \omega < \omega_{\rm sj}  \\
    \mathcal{S}^{\rm jit}_{\omega}[\gamma,\delta B,\Psi_{\rm slab}(k),\theta] \quad{\rm for} \quad \omega > \omega_{\rm sj} \\
    \end{cases}
\end{multline*}
where the transition frequency is:
\begin{align}
    \omega_{\rm sj}  \simeq \frac{2\gamma^2k_\lambda c}{1+\gamma^2\theta^2}
\end{align}

\begin{figure*}[h!]
    \begin{centering}
\includegraphics[ width=0.99\textwidth]{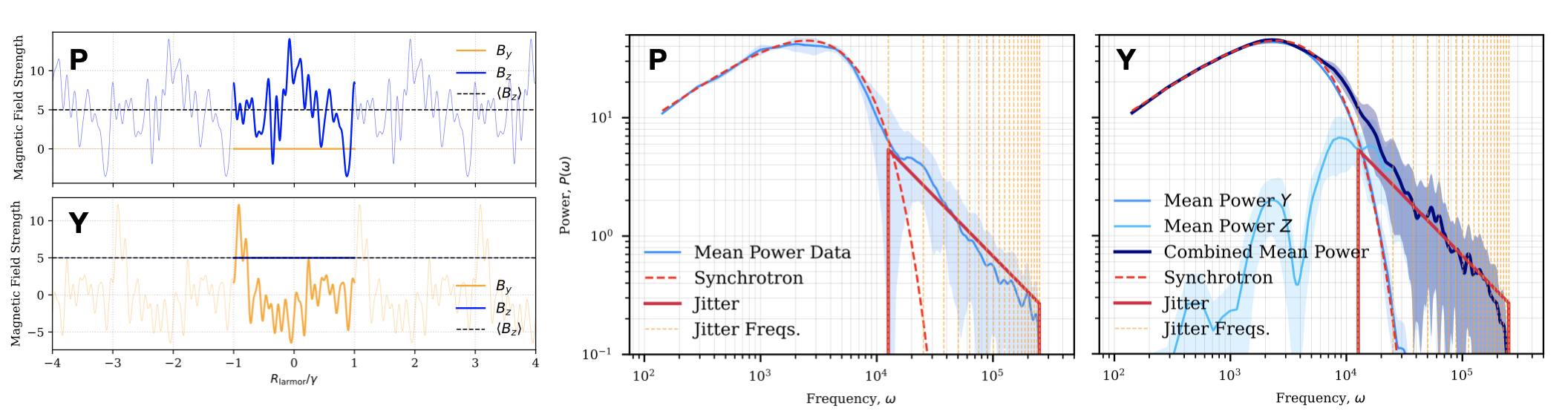}
\end{centering}
\caption{\textit{Left panel:} the magnetic field along the particle trajectory, for a typical simulation. The highlighted portion is the one in the emission window. Upper plot for pure-pitch (P), lower plot for pure-yaw (Y). \textit{Center and right panel:} the average power spectrum, seen by an observer in the same plane of the unperturbed orbit, computed from Liénard–Wiechert potentials, over 50 realization of the magnetic field and particle trajectory, with the synchrotron and jitter components, for pure pitch (P) and yaw (Y).
For the pure-yaw case, we plot the total power as well as the power in the two orthogonal modes of the electric field. The discrete jitter frequencies, corresponding to our discrete modes are also shown. Shaded regions represent the $1\sigma$ uncertainty.  All plots are in units: $e=c=m=1$, and the particle has $\gamma=20$.  }
    \label{fig:sj_spectrum}
\end{figure*}

We have tested this emission formula with a direct calculation of the observed power spectrum, computed using the retarded Liénard–Wiechert potentials from the true orbit of the particle, for an observer located along the $x$-axis.
We selected two different regimes for magnetic perturbations: \textit{pure-pitch}  such that the particle orbit remains confined to the same plane of the unperturbed orbit ($\delta\boldsymbol{B} \parallel \boldsymbol{B}$); and \textit{pure-yaw} where the jittering is orthogonal to the unperturbed orbital plane ($\delta\boldsymbol{B} \perp \boldsymbol{B}$). Due to the symmetry of the configuration, in the pure-pitch case the radiation electric field will be in the $xy$ plane, while the pure-yaw case will have both components in and perpendicular to that plane. In both cases, magnetic perturbation as a function of position $x$ (the longitudinal coordinate in the emission window) were introduced using a discrete spectrum with a power-law index $p=1$:
\begin{multline*}
    \delta B (x) = B_{z} \left[ \sum_{i=1}^{20} A_i \cos(i k_\lambda x + \phi_i) \right]
    \\
    {\rm with}\quad 
    A_i = i^{-p/2}\left[\sum_{j=1}^{20} j^{-p}\right]^{-1/2}
\end{multline*}
and random phases $\phi_i$. In Fig.~\ref{fig:sj_spectrum} we show the results. Assuming that magnetic  turbulence has a common universal spectrum $\Psi(k)$, if $\mathcal{P}[B,\delta B]$ describes the probability distribution function for the mean and rms magnetic field in a given region of the source, the total emissivity will be given by:
\begin{equation}
    \mathcal{S}_\omega^{\rm sj\;tot}[\gamma,\theta] = \int \mathcal{S}_\omega^{\rm sj}[\gamma,B,\delta B, \Psi_{\rm slab}(k),\theta] \mathcal{P}[B,\delta B ]dB d\delta B
\end{equation}

\bibliography{Total,References}

\end{document}